\documentclass[reprint,
superscriptaddress,
 amsmath,amssymb,
notitlepage,
aps,
prl,
10pt, 
tightenlines
]{revtex4-2}

\usepackage[utf8]{inputenc}
\usepackage[caption=false]{subfig}
\usepackage{graphicx}
\usepackage{dcolumn}
\usepackage{bm}
\usepackage{xcolor}
\usepackage{verbatim}
\usepackage{tabularx}
\usepackage{bbold}
\usepackage{float}
\usepackage{bm}
\usepackage{bbm}
\usepackage[colorlinks=True, citecolor=blue, urlcolor=blue]{hyperref}
\usepackage{booktabs}
\usepackage{natbib}

\usepackage{tikz}
\usetikzlibrary{quantikz}

\definecolor{crimson}{rgb}{.8, 0, 0}

\newcommand{\bigO}[1]{\mathcal{O}\left( #1 \right)}
\newcommand{\phdag}{\phantom{\dagger}}

\begin{document}
\title{The quantum adiabatic algorithm suppresses the proliferation of errors}
\author{Benjamin~F.~Schiffer}
\email[Corresponding author: ]{Benjamin.Schiffer@mpq.mpg.de}
\author{Adrian~Franco~Rubio}
\affiliation{Max-Planck-Institut f\"ur Quantenoptik, Hans-Kopfermann-Str.~1, D-85748~Garching, Germany}%
\affiliation{Munich Center for Quantum Science and Technology (MCQST), Schellingstr.~4, D-80799~Munich, Germany}%
\author{Rahul~Trivedi}
\affiliation{Electrical and Computer Engineering, University of Washington, Seattle, WA~98195, USA}
\affiliation{Max-Planck-Institut f\"ur Quantenoptik, Hans-Kopfermann-Str.~1, D-85748~Garching, Germany}%
\affiliation{Munich Center for Quantum Science and Technology (MCQST), Schellingstr.~4, D-80799~Munich, Germany}%
\author{J.~Ignacio~Cirac}
\affiliation{Max-Planck-Institut f\"ur Quantenoptik, Hans-Kopfermann-Str.~1, D-85748~Garching, Germany}%
\affiliation{Munich Center for Quantum Science and Technology (MCQST), Schellingstr.~4, D-80799~Munich, Germany}%

\date{\today}

\begin{abstract}
The propagation of errors severely compromises the reliability of quantum computations. The quantum adiabatic algorithm is a physically motivated method to prepare ground states of classical and quantum Hamiltonians. Here, we analyze the proliferation of a single error event in the adiabatic algorithm. We give numerical evidence using tensor network methods that the intrinsic properties of adiabatic processes effectively constrain the amplification of errors during the evolution for geometrically local Hamiltonians. Our findings indicate that low energy states could remain attainable even in the presence of a single error event, which contrasts with results for error propagation in typical quantum circuits.

\end{abstract}

\maketitle

\emph{Introduction.}---Current quantum hardware is increasingly becoming competitive with classical computational capabilities~\cite{Morvan2023Phase, King2024Computational}. A key bottleneck towards practical applications are hardware errors and their proliferation during the computation. Interestingly though, the susceptibility towards hardware noise can vary strongly depending on the specific algorithm used and the type of error. The adiabatic algorithm~\cite{Farhi2000Quantum} is a paradigmatic model of quantum computation, where the proliferation of errors remains unexplored.

Hardware noise in  quantum computers has severe implications for the ability to perform useful quantum computations. It was realized early on that in the presence of depolarizing-like noise, the computational state approaches the maximally mixed state exponentially fast~\cite{Aharonov1996Limitations}. This was found to put stringent limitations on the ability to solve classical optimization problems on noisy quantum computers with an advantage compared to classical computers using variational or adiabatic algorithms~\cite{StilckFranca2021Limitations,DePalma2023Limitations}. 
The situation can be even worse due to the propagation of errors since even a single error in the middle of the circuit can proliferate during the quantum computation and render the computation useless.
This phenomenon was recently studied in the random circuit model~\cite{GonzalezGarcia2022Error, Mishra2024Classically}.
In practice, though, gates are not drawn from a random distribution, but depend on the particular algorithm.

Analogue quantum simulators are a particularly interesting class of quantum devices that probe physics that is often not accessible by classical methods~\cite{Ebadi2020Quantum, Scholl2021Quantum, Wei_2022}. The prospect of current quantum simulators providing a quantum computational advantage has been recently studied~\cite{Flannigan2022Propagation, Daley2022Practical}, and there is evidence that quantum simulators can solve certain problems in physics stably and robustly to errors for intensive observables~\cite{Trivedi2022Quantum, Kashyap2024Accuracy}. 
In this context, the quantum adiabatic algorithm plays a very relevant role. The algorithm relies on smoothly interpolating from a trivial initial Hamiltonian $H_0$ towards a target Hamiltonian $H_T$ in order to prepare its ground state. If the ground state during the evolution is separated by a spectral gap from higher eigenstates, the adiabatic theorem guarantees that the ground state of the target Hamiltonian can be prepared with an evolution time that scales inverse polynomially with the gap~\cite{Kato1950Adiabatic, Messiah1962Quantum, Farhi2000Quantum, Jansen2007Bounds, Amin2009Consistency, Albash2018Adiabatica}. 
Interestingly, the adiabatic algorithm has some robustness to hardware noise, in particular against decoherence and control errors~\cite{Childs2001Robustnessa, Roland2005Noise, Aaberg2005Robustness, Albash2015Decoherence}.

\begin{figure}[b]
    \includegraphics[width=1.\linewidth]{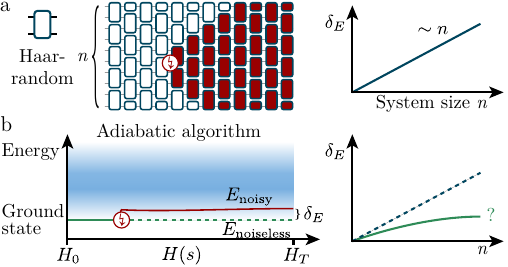}
    \caption{\textbf{(a)} A single depolarized qubit in a circuit of Haar-random gates can render the whole computation unreliable due to the proliferation of the error. If a deep quantum circuit provides a solution to a problem, e.g.~the minimization of a cost function $\mathcal{C}$, the noisy circuit propagates the error and prepares a state with an extensive error $\delta_E = |\mathcal{C}(\text{no error}) - \mathcal{C}(\text{error})|$. 
    \textbf{(b)} The quantum adiabatic algorithm follows a different paradigm, where the ground state of an initial Hamiltonian is slowly transformed into a target state. The cartoon shows a low-energy state as the output of an adiabatic sweep in the presence of a single error.  Darker shades of blue in the middle of the spectrum indicate a higher density of states, typical for local Hamiltonians. We give evidence that error proliferation in the adiabatic algorithm is constrained, yielding a subextensive growth of the energy error.}
    \label{fig:Fig1}
\end{figure}

In this Letter, we provide an analysis of the proliferation of errors in the quantum adiabatic algorithm for local Hamiltonians. As a figure of merit, we consider the energy error $\delta_E$, defined as the difference between the observed energy after a noisy evolution and the noiseless evolution~\footnote{For classical Hamiltonians encoding combinatorial optimization problems the energy error is typically equivalent to the approximation ratio.}. 
In our analysis, we focus on a single error as with random circuits this can already have catastrophic consequences for sufficiently deep circuits, on average resulting in a state that is locally indistinguishable from the maximally mixed state. Thus, a single depolarized qubit results in an extensive energy error when interpreting the computation as a variational quantum circuit [Fig.~\ref{fig:Fig1}(a)]. 
We then give numerical evidence that for the adiabatic algorithm with a local Hamiltonian, the energy error does not scale extensively as in random circuits, but at a lower rate [cf. Fig.~\ref{fig:Fig1}(b,c)]. For an adiabatic path within the integrable parameter range of the Hamiltonian, we explain these findings with a proof showing that a single noise event yields an energy error that is bounded by a constant, independent of system size. 
Our analysis combines analytical tools with large scale simulations using matrix product states (MPS) of up to 100 spins. The results give clear evidence that error propagation is constrained in the adiabatic algorithm.

\emph{Single error model.}---To study the proliferation of errors, we consider a single error during the quantum computation. 
A single erroneous qubit can have severe consequences for the output of the computation. When a quantum circuit creates entanglement, the error propagates through the circuit and is bounded by a causal light cone. An error is generically expected to impact every qubit in the light cone. This can be made precise for random circuits, which are good models for variational quantum algorithms~\cite{Cerezo2021Variational, Bharti2022Noisy, Anschuetz2022Quantum}. 
For these random circuits, the average state after a single depolarized qubit in a deep circuit is locally indistinguishable from the completely mixed state $\rho = \mathbb{1}^{\otimes n}/2^n$. This builds on previous results~\cite{GonzalezGarcia2022Error}, and we include the proof for completeness in the Supplement~\cite{Supp}. If the noiseless circuit prepares the ground state of a target Hamiltonian, the noisy circuit prepares a state that has an energy difference $\delta_E \sim n$ with the target state. 

\begin{figure}[t]
    \includegraphics[width=1.0\linewidth]{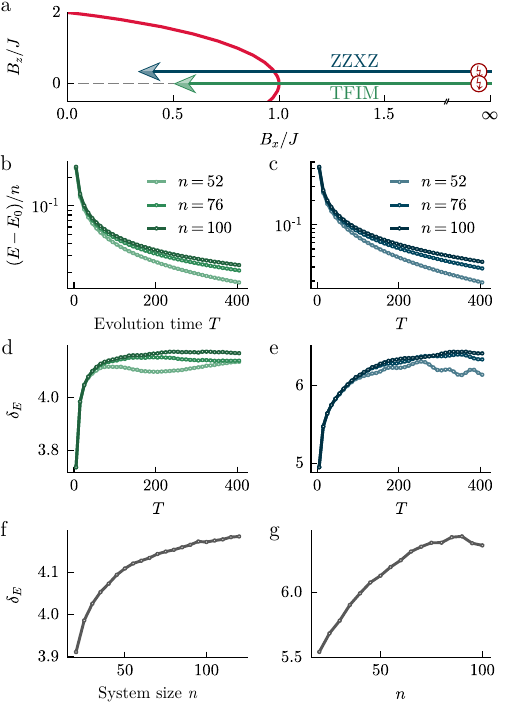}
    \caption{
    \textbf{(a)} Phase diagram of the mixed field Ising model (ZZXZ) showing two sweeps of the adiabatic algorithm. The sweep along $B_z=0$ considers the transverse field Ising model (TFIM).
    \textbf{(b)} Energy density above the ground state for a noiseless, linear adiabatic sweep in the TFIM, converging asymptotically to the ground state energy. 
    \textbf{(c)} Same data for the ZZXZ model, corresponding to the blue arrow in the phase diagram. 
    \textbf{(d)} The excess energy $\delta_E$ in the single error model plateauing for the TFIM as a function of evolution time $T$. 
    \textbf{(e)} Same data for the ZZXZ model in an adiabatic path from $H_\textsc{ZZXZ}(J=B_z=0, B_x=1)$ to $H_\textsc{ZZXZ}(J=3, B_x=B_z=1)$. 
    \textbf{(f)} The excess energy $\delta_E$ increases sublinear as a function of system size $n$ for the TFIM. The evolution time is scaled as $T=n^2/40$ to remain adiabatic. 
    \textbf{(g)} The ZZXZ model also shows a sublinear increase of $\delta_E$ with $n$, implying a constrained proliferation of errors ($T=n^2/20$).}
    \label{fig:Fig2}
\end{figure}

\begin{figure*}[t!]
    \includegraphics[width=1.0\linewidth]{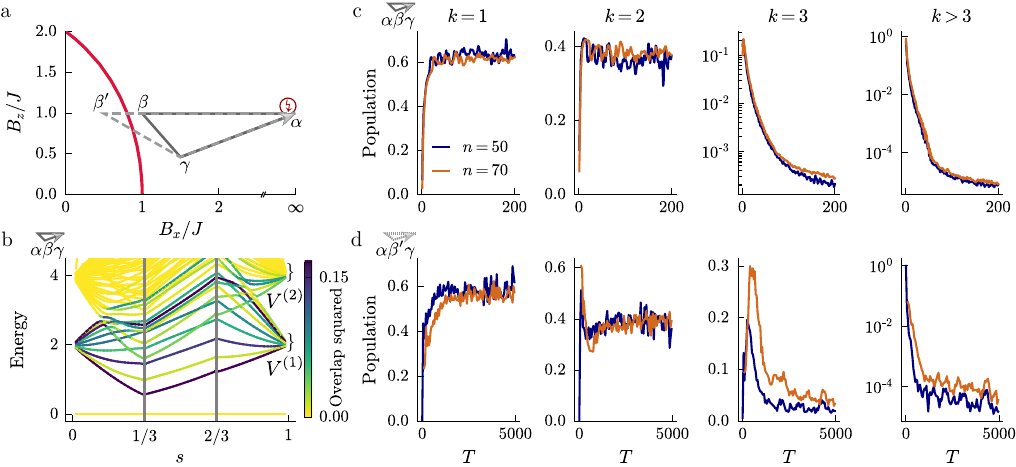}
    \caption{\textbf{(a)} Adiabatic loop in the disordered phase of the mixed field Ising model ($\triangle\alpha\beta\gamma$, solid line) and another loop that crosses the phase transition twice ($\triangle\alpha\beta'\gamma$, dashed). 
    \textbf{(b)} Simulated dynamics (exact diagonalization) for an adiabatic loop in the disordered phase with an noisy qubit in the beginning. Population is transferred to low-energy excited states, but remains in the first two excitations spaces. 
    \textbf{(c)} Population as a function of total evolution time $T$ (per leg) for the $k=1,2,3$ excitation space and everything above. The tensor network simulations are for the same loop as in (b) and confirm that the error does not proliferate to higher excited states for deep circuits and large system sizes. 
    \textbf{(d)} Similar simulations for the loop crossing the phase transition showing a strong similarity to the noncritical loop. For the critical loop, the dynamics are becoming approximately adiabatic (50 spins: 99.8\% ground state fidelity at $T=5000$; 98.6\% for 70 spins). We observe small leakage into the third excitation space in this regime. Note that the populations feature an oscillation for small $T$, induced by a phase shift which depends on $T$ through the sweep velocity at the avoided-level crossings, additional data in~\cite{Supp}.}
    \label{fig:Fig3}
\end{figure*}

\emph{Numerical simulations of adiabatic algorithms.}---The quantum adiabatic algorithm is of very different nature than random circuits. It is then natural to ask about error proliferation in the quantum adiabatic algorithm using the single error model.
We limit our investigations mostly to the antiferromagnetic ZZXZ model 
\begin{align}
    H_\textsc{ZZXZ} = J \sum_i \sigma^z_i \sigma^z_{i+1} + B_x \sum_i \sigma^x_i + B_z \sum_i \sigma^z_i,    
\end{align}
and perform numerical simulations using the TEBD algorithm~\cite{Vidal2004Efficient}.
The ZZXZ model is non-integrable when both field strengths $B_x$, $B_z$ are finite, and features a second order phase transition from the disordered phase into the antiferromagnetic phase [Fig.~\ref{fig:Fig2}(a)]. The adiabatic gap vanishes as \mbox{$\Delta_\text{ad} \sim 1/n$~\cite{Ovchinnikov2003Antiferromagnetic}}.
For both trajectories in the phase diagram, we consider linear sweep schedules with a total evolution time $T$ such that the noiseless evolution is approximately adiabatic for large $T$ [Fig.~\ref{fig:Fig2}(b,c)]. 
Next, we include a single error early in the dynamics and show the resulting excess energy $\delta_E$ in Fig.~\ref{fig:Fig2}(d,e). We consider the unitary error operator $\sigma^y$ on a qubit in the middle of the chain early in the dynamics, as indicated in Fig.~\ref{fig:Fig2}(a)~\footnote{A depolarizing noise channel on the $i$th qubit $\mathcal{N}^{(d)}_i$ is equivalent to stochastic Pauli noise as 
\begin{align}
    \rho &\rightarrow \mathcal{N}^{(d)}_i\rho[\mathcal{N}^{(d)}_i]^\dag \\
   &= (1-p)\rho + \frac{p}{3}(\sigma^x_i\rho\sigma^x_i + \sigma^y_i\rho\sigma^y_i + \sigma^z_i\rho\sigma^z_i).
\end{align}
where $p/3$ is the probability to trigger one of the Pauli matrices.}. We observe that $\delta_E$ in the adiabatic limit does not grow larger for longer evolution times, i.e.~deeper circuits.
When increasing the system size, the evolution times needs to be rescaled with $T\sim n^2$ to remain adiabatic~\cite{Jansen2007Bounds}. The numerical results in Fig.~\ref{fig:Fig2}(f,g) show  that the excess energy as a function of system size clearly increases slower than linear, implying that every qubit is impacted by noise more weakly than it would be in the case of a random circuit.

For a particular class of models, we prove the constrained proliferation of errors in the adiabatic algorithm. We consider fermionic Gaussian Hamiltonians (FGH) that are of the form $H_\textsc{FGH}=\sum_{i,j} h_{ij} c_i c_j$, where the $c_i$ are Majorana operators fulfilling fermionic anticommutation relations.
For unitary dynamics generated by Hamiltonians of this class, we show that the excess energy $\delta_E$ due to a single error is bounded by a constant, independent of the system size. The local noise operator can be expressed as a polynomial in the fermionic modes on those sites where the noise operator has support. The unitary evolution then merely dresses the excitation during the adiabatic algorithm without increasing the degree of the polynomial. Hence, the energy error after the dynamics is still independent of the system size; we include a formal version of this proposition in the Supplement~\cite{Supp}.
The adiabatic sweep in Fig.~\ref{fig:Fig2} along the $B_z=0$ axis implements the transverse field Ising model, which can be brought into the form of a $H_\textsc{FGH}$ through the Jordan-Wigner transformation~\cite{Jordan1993Ueber}. 

\emph{Low energy excitations.}---Motivated by the analysis of Gaussian models, we continue to investigate non-integrable models by analyzing excitations on top of the ground state.
For convenience, we consider, as starting and ending states for the adiabatic algorithm, points in the phase diagram where the spectrum is given by product states.
For these states we can easily define the $k$-excitation space $V^{(k)}_{\ket{p}}$, which is the space spanned by all states obtained by applying a total of $k$ different single-site excitations to the state $\ket{p}$. Concretely, for an operator $\sigma_i^{\alpha}$ acting on the $i$th spin, the 1-excitation space is $V^{(1)}_{\ket{p}} = \text{span}[\sigma_1^{\alpha}\ket{p}, \dots, \sigma_n^{\alpha} \ket{p}]$.

The initial state of the adiabatic algorithm is a product state and we cycle through the phase diagram back to the starting point, as shown in Fig.~\ref{fig:Fig3}(a). First, we consider a loop ($\triangle\alpha\beta\gamma$) that remains within the disordered phase of the Ising model. In Fig.~\ref{fig:Fig3}(b), we show a spectrum along this loop for a small system of 10 spins. The noise gate $\sigma^y$ is applied on a qubit in the center of the chain at the beginning of the evolution. The populated eigenstates are clearly restricted to low-energy states. At the end of the loop the population is contained within the first two bands of eigenstates above the ground state. This holds also when scaling up the system size to up to 70 spins in Fig.~\ref{fig:Fig3}(c). We can quantify that the first and second excitation spaces are populated with approximately $40\%$ and $60\%$, respectively, in the limit of a slow sweep. 
Next, in Fig.~\ref{fig:Fig3}(a,d), we show the behavior of a loop ($\triangle\alpha\beta'\gamma$) that crosses the critical region. 
We observe that the containment of errors to the low energy subspace is still valid in this case. Concretely, when the dynamics are approximately adiabatic at large $T$, the state occupies nearly exclusively the lowest two energy subspaces above the ground state. A small share of the population of 2\% is found in the $k=3$ subspace for 50 spins, and 3\% for 70 spins, respectively. The overlap squared with higher energy states is of the order of $10^{-4}$ and decays further when the dynamics are even slower.

As a remark regarding the spectrum in Fig.~\ref{fig:Fig3}(b), we note that even though there are in principle additional lines crossing with populated eigenstates, the population is contained to low energy states. This is because the proliferation of errors in an adiabatic algorithm also depends on the matrix elements of the Hamiltonian $V_{ij} = \langle \phi_i(s) | \partial H(s)/\partial s | \phi_j(s) \rangle$ that govern transitions between the eigenstates (we define $H(s) \ket{\phi_i(s)} = \lambda_i(s) \ket{\phi_i(s)}$). Symmetries of the Hamiltonian render some $V_{ij}$ zero and prevent population from being transferred to higher energies. In particular, we consider the translation operator $T$ that shifts the system by one site. The eigenvalues of $T$ define the momentum $q$ as $T\ket{\varphi}=\exp(i\pi q) \ket{\varphi}$. For translationally invariant system $[H(s),T]=0$ for all $s$ and the eigenstates $\{\ket{\phi_i(s)}\}_i$ have a well-defined momentum $q$. Different momentum sectors are effectively decoupled, creating a finite gap between the first eigenstate of the $q=0$ sector and higher eigenstates. The models in the numerical benchmark have open boundary conditions, hence, they are approximately translationally invariant for large system sizes.

\begin{figure}[t]
    \includegraphics[width=1.0\linewidth]{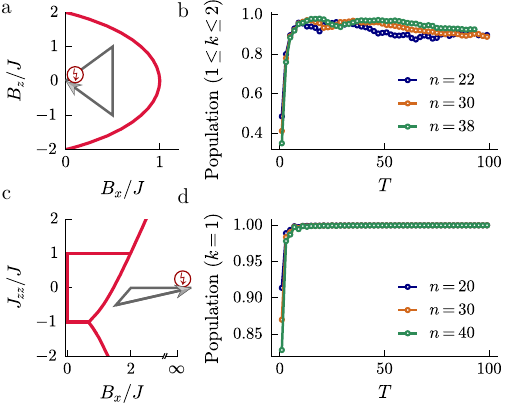}
    \caption{\textbf{(a)} Adiabatic loop within the antiferromagnetic phase of the mixed field Ising model.
    \textbf{(b)} The noisy evolution is largely contained in the first two excitation spaces. We note that error proliferation is dependent on whether the error occurs on an even or odd site of a Néel state. The data here is for the benign case and additional data is included in~\cite{Supp}.
    \textbf{(c)} Adiabatic loop in the disordered phase of a non-integrable Heisenberg model. Critical lines are again shown in red, and the disordered phase is found for large, positive $B_x/J$. 
    \textbf{(d)} For the loop in the disorder phase of this model, the population is fully contained in the first excitation space.}
    \label{fig:Fig4}
\end{figure}

\emph{Further studies on non-integrable models.}---Most of the numerical experiments up to this point targeted the disordered phase of the mixed field Ising model. To show that the robustness of the adiabatic algorithm appears to be a more general feature, we also consider other phases. First, we consider the antiferromagnetic phase, where we start in the ground state of the Ising model with zero field strengths [Fig.~\ref{fig:Fig4}(a)]. At this point, the ground state is again a product state and the excitation space is again defined as the first eigenenergy space above the ground state. We consider an adiabatic loop back to the initial point and observe a high population in the first excitation space that is approximately independent of system size [Fig.~\ref{fig:Fig4}(b)].  Note that, in the antiferromagnetic phase, whether an odd or even spin is subjected to noise influences how much the energy of the system increases immediately due to the noise event, influencing the excess energy at the end of the computation~\cite{Supp}.

Additionally, we consider the disordered phase of a non-integrable Heisenberg model $H_\text{Heis} = 1/2 \sum_i (J\sigma^x_i \sigma^x_{i+1} + J\sigma^y_i \sigma^y_{i+1} + J_{z} \sigma^z_i \sigma^z_{i+1}) + B_x \sum_i \sigma^x_i$~\cite{Dmitriev2002One}. Note that the magnetic field is not aligned with the $z$ direction of the spins, which breaks Bethe-integrability.
We consider an adiabatic loop in the disordered phase and observe near-perfect containment in the first excitation space [Fig.~\ref{fig:Fig4}(c,d)], which can be explained by few energy lines crossing during the evolution and only small coupling between these eigenenergies.

\newpage
\emph{Discussion and outlook.}---The study of noise in quantum systems is a complex matter that remains very relevant while the hardware is being built up to practical quantum error correction. In this Letter, we provide a novel perspective on error propagation in a physically inspired quantum algorithm. We study the single error model and show that the adiabatic algorithm behaves strikingly different than a random circuit. We consider the excess energy as a figure of merit and show that the excess energy for integrable, free fermionic models can be bounded by a constant. Interestingly, the non-integrable models in our benchmark also show a significantly constrained proliferation of errors. We remark that in the presence of depolarizing noise, there are no-go results that cannot be avoided~\cite{Aharonov1996Limitations, Franca2020Limitations}. If this is not the case, however, our results could provide evidence that the adiabatic algorithm works well in practice~\cite{Ebadi2022Quantum}. 

We show that a single error can be understood as a dressed excitation on top of the ground state for fermionic Gaussian Hamiltonians. Our results suggest that this behavior also applies approximately for integrable models. We leave for future studies to make the observation rigorous. 
Also, we have only considered a single error, and expect that if there are few errors, e.g.~that as long as the errors are not extensive, the result would hold since they would not interact with each other. However, this is an interesting question that deserves to be studied further.

A key feature of the quantum adiabatic algorithm for physical models is that it does not create volume law entanglement, which is desirable in the presence of hardware noise as entanglement accelerates the proliferation of errors~\cite{GonzalezGarcia2022Error}. Further research could explore if there exists a deeper connection behind the constrained proliferation of errors and the limited entanglement during the algorithm.
Another possible directions is to relate the robustness of the adiabatic algorithm to an approximate conservation of energy in the presence of a slowly changing Hamiltonian. 

More generally, our study highlights that the choice of the algorithm can have substantial consequences for how strongly noise affects the results of a quantum computation. Our results motivate the design of quantum algorithms that are particularly resilient to noise, possibly by relying on similar physical mechanisms as in the adiabatic algorithm.

\begin{acknowledgments}
The authors acknowledge Guillermo González-García for insightful discussions. This research is part of the Munich Quantum Valley, which is supported by the Bavarian state government with funds from the Hightech Agenda Bayern Plus. We acknowledge funding from the German Federal Ministry of Education and Research (BMBF) through EQUAHUMO (Grant No.~13N16066) within the funding program quantum technologies---from basic research to market. A.F.R.~acknowledges support by the Alexander von Humboldt Foundation through a postdoctoral fellowship.
\end{acknowledgments}

\bibliography{biblio}
\clearpage
\appendix

\let\oldsection\section
\renewcommand{\section}[1]{\oldsection*{\arabic{section}. #1}\stepcounter{section}}
\setcounter{section}{1} 

\begin{center}
\textbf{\large Supplemental Material: \\The quantum adiabatic algorithm suppresses the proliferation of errors}
\end{center}

\setcounter{equation}{0}
\setcounter{figure}{0}
\setcounter{table}{0}
\setcounter{page}{1}
\makeatletter
\renewcommand{\theequation}{S\arabic{equation}}
\renewcommand{\thefigure}{S\arabic{figure}}
\renewcommand{\bibnumfmt}[1]{[S#1]}

\section{Single error in random circuits}

We consider the propagation of errors in a quantum circuit where every gate is drawn from a Haar-random distribution. This scenario was analyzed in Ref.~\cite{GonzalezGarcia2022Error}. We build upon their results to show that the energy error in the final state after a single depolarized qubit in the circuit scales extensively with the system size $n$. Without loss of generality, we focus on the 1D case.

Importantly, we assume a brick-wall layer structure of the circuit and gates that act on two neighboring qubits simultaneously. 
The twirl of a two-qubit quantum channel $\mathcal{M}$ over the Haar distribution is again a two-qubit depolarizing channel as
\begin{align}
    \mathcal{E}_d (\rho) = \int_\mathcal{U} d\mathcal{U} [\mathcal{U}^\dag \mathcal{M} \mathcal{U}](\rho)
\end{align}
We restate the three possible outcomes:
\begin{enumerate}
    \item If no error occurs [$\mathcal{M}(\rho)=\rho$], then $\mathcal{E}_d (\rho) = \rho$;
    \item If two errors occur [$\mathcal{M}(\rho)=\text{Tr}(\rho)\mathbb{1}^{\otimes 2}/4$], then $\mathcal{E}_d(\rho) = \text{Tr}(\rho) \mathbb{1}^{\otimes 2}/4$;
    \item If an error occurs in one of the two qubits [e.g.~on the first $\mathcal{M}(\rho)=\mathbb{1}/2 \otimes \text{Tr}_1(\rho)$], then 
    \begin{align}
        \mathcal{E}_d(\rho) = \frac{1}{5}\rho + \frac{4}{5} \text{Tr}(\rho) \frac{\mathbb{1}^{\otimes 2}}{4}.
    \end{align}
\end{enumerate}
The error propagation can be modeled by a Markov chain as follows. We consider a string of zeros and ones, such that a ``0'' represents a noiseless qubit and a ``1'' represents a noisy qubit. As the state $X$ of the Markov chain we consider the number of ones in the string. Then, the state $X=0$ represents a noiseless system, and a system with a single depolarized qubit is in the $X=1$ state. Lemma~2 in Ref.~\cite{GonzalezGarcia2022Error} states the transition matrix for the states $X$. We summarize in words: From $X=1$ the state goes to $X=0$ with probability 1/5, and with probability 4/5 the error propagates ($X=2$). The steady states of the chain are the $X=0$ and the $X=n$ states. The probability of reaching $X=0$ after $t$ applications of the Markov chain can be bounded (Lemma~3~in Ref.~\cite{GonzalezGarcia2022Error}):
\begin{align}
    \text{Pr}(X_t=0|X_t=1) \leq \frac{1}{4} \quad\forall t.
\end{align}
The average number of depolarized qubits at time t $\langle q(t) \rangle$ is then given as 
\begin{align}
\langle q(t) \rangle \approx \frac{3}{4} \text{min}\left(\frac{6t}{5}, n\right),
\end{align}
which is approximate~\cite{GonzalezGarcia2022Error}. 
This implies, on average, with a probability of more than 75\%, a single depolarized qubit is equivalent as the completely mixed state $\rho_m = \mathbb{1}^{\otimes n}/2^n$. 

The final step in our argument is that for a local Hamiltonian $H$, the difference between the ground state energy $E_0$ and the highest eigenenergy $E_{2^n-1}$ is extensive: $E_{2^n-1} - E_0 \sim n$. Without loss of generality we can assume $H$ to be traceless, such that the energy of the completely mixed state is $E_m = \text{Tr}(H \rho_m) = 0$. Hence, a single depolarized qubit propagates the error in random circuits such that, on average, the energy of the prepared states $E_\text{noisy}$ under the Hamiltonian $H$ differs from the target state energy $E_0$ as $\delta_E = E_\text{noisy} - E_0 \sim n$.

\section{No error propagation in Gaussian systems}

Here we present the argument that Gaussian fermionic systems do not give rise to error propagation scaling with the system size. The setting is as follows: we consider a lattice system defined by local fermionic modes $a_j^{\phdag}, a^\dagger_j$, an initial Gaussian state $\ket{\text{init}}$ (for instance the vacuum state $a_j\ket{\text{init}}=0~~\forall j$) and a  fermionic Gaussian Hamiltonian,
\begin{equation}
    H = \sum_{i,j}{h_{ij}a^\dagger_ia_j^{\phdag} + \tilde h_{ij} a^\dagger_ia_j^{\dagger} + \text{H.c.}},
\end{equation}
whose ground state $\ket{\text{GS}}$ is therefore a Gaussian state. In fact, any Hamiltonian of this form can be diagonalized by a change of basis of the modes, 
\begin{equation}
    H = \sum_{k}{\omega_k  b^\dagger_k b^{\phdag}_k},
\end{equation}
where $b^{\phdag}_k, b^\dag_k$ are linear combinations of the original modes $a_j^{\phdag}, a^\dagger_j$. We will assume that the single particle energies are bounded, $\max_k{\omega_k}=\bigO{1}$, as is the case in most physically relevant models, e.g.~translationally invariant systems, where $\omega_k$ corresponds to the dispersion relation.

Now assume there exists a Gaussian unitary $U$, representing our algorithm, such that $U \ket{\text{init}} = \ket{\text{GS}}$. A unitary is Gaussian if it preserves the family of Gaussian states: such maps amount once again to linear transformations of the fermionic modes. The precise nature of $U$ is not important for the argument: it may have been obtained from a variational circuit $U=\prod u_j$, or as a result of adiabatic evolution $U=\mathcal{T}\exp(-i\int_0^s ds H(s))$ along a path of Gaussian Hamiltonians $H(s)$. 

Lastly, consider a local noise operator $\mathcal{N}$ supported on $s=\bigO{1}$ sites (without loss of generality, and for notational convenience, we take them to be the sites labeled $1,\ldots,s$). $\mathcal{N}$ can therefore be expressed as a polynomial in the fermionic modes on these sites
\begin{equation}
    \mathcal{N} = \sum^{2s}_{n=1}{\sum_{i_1,\ldots,i_n}{r^{(n)}_{i_1,\ldots,i_n}}\alpha_{i_1}\ldots\alpha_{i_n}}
\end{equation}
where $\alpha=(a^{\phdag}_1, a^{\phdag}_2, \ldots, a^{\phdag}_s, a^{\dag}_1, a^{\dag}_2, \ldots, a^{\dag}_s)$ is the vector of the creation-annihilation operators on sites $1,\ldots, s$ and the $r^{(n)}$ are the coefficients of each monomial. Importantly, this polynomial will be of bounded degree $2s$, as there are only $s$ fermionic modes in its support.

If the noise event takes place at the beginning of our protocol, the final state will be
\begin{equation}
    U \mathcal{N} \ket{\text{init}} =  U \mathcal{N} U^\dag U \ket{\text{init}} = U \mathcal{N} U^\dag\ket{\text{GS}}.
\end{equation}
We want to argue that the energy difference with the ground state is bounded for this state. The key to the argument is that $U \mathcal{N} U^\dag$ is once again a polynomial of bounded degree $2s$ in $a_j^{\phdag}, a^\dagger_j$, since $U$ merely implements a linear transformation of the modes. Furthermore, this property is preserved when changing to the eigenbasis $b_j^{\phdag}, b^\dagger_j$, which amounts to another linear transformation. Thus, the state prepared by the noisy protocol is of the form
\begin{equation}
    \left(r^{(0)}+r^{(1)}_{k}b^\dag_k+r^{(2)}_{kk'}b^\dag_kb^\dag_{k'}+\ldots\right)\ket{\text{GS}},
\end{equation}
which is a superposition of states with at most $2s$ excitations and thus bounded energy difference with the ground state,
\begin{equation}
\delta_E\leq2s\max_k{\omega_k}=\bigO{1}.
\end{equation}

\section{Bound on populated states for a local error}
For completeness, we briefly state results that bound the overlap with high energy states when the ground state is perturbed locally. 

Consider a state $\ket{\psi(s)}$ during the adiabatic algorithm before the noise event occurs. The Hamiltonian $H(s)$ generating the dynamics is at all times assumed to be (geometrically) local. Let $\lambda(s)$ be the energy of the state during the protocol. Then, an arbitrary error on a single qubit at time $s^*$ can increase the energy at most by some $c$ which depends on the locality and norm of the Hamiltonian, but is independent of system size. Concentration bounds for quantum states exist that bound the overlap of this state with high energy states: 
For a state on a $D$-dimensional lattice, with correlation length $\sigma$, and $m$ local terms in the Hamiltonian, the overlap $\sqrt{p_f}$ with the eigenspace of energy $f$ is bounded as
\begin{align}
    \sqrt{p_f} \leq \exp\left\{\frac{-\left[(\lambda(s^*)+c - f)^2 \sigma \right]^{1/(D+1)}} {m^{1/(D+1)}D\sigma}\right\}
\end{align}
when $|\lambda(s^*)+c - f|>2^D\sqrt{m\sigma}$~\cite{Anshu2016Concentration}.
For one-dimensional gapped systems, the overlap with higher excited states then decays exponentially. 

\section{Details on the numerical simulations}
We implement the numerical simulations using the TEBD algorithm~\cite{Vidal2004Efficient} and ITensor~\cite{Fishman_2022}. We do not set a maximal bond dimension during the dynamics, and only use a precision cutoff of $\epsilon=10^{-8}$ for truncating bonds when updating tensors in the simulations. As the dynamics after a single noise event do not follow the ground state trajectory, a strict area law does not apply. These violations of the area law then lead to a larger bond dimensions required in these simulations compared with the noiseless case. Ground states are computed using the DMRG algorithm~\cite{Schollwoeck2011density} up to numerical precision. 

The single noise event is modeled by the application of a $\sigma^y_i$ operator acting on the $i$th qubit. This error can be related to depolarizing noise, which corresponds to the application of a random Pauli matrix via quantum trajectory method~\cite{Moelmer1993Monte}. Closed system dynamics with all three Pauli operators on the $i$th qubit suffice for the simulation of the depolarized qubit. Note that early in the sweep, the commutator $[\sigma^y_i, H_\textsc{ZZXZ}(s=0)]$ is the largest for the three Pauli operators. Therefore, we focus on $\sigma^y_i$, which can be related to a bound on the excess energy due to a depolarized qubit. See also Fig.~21(b) in~\cite{Schiffer_2022} where the different Pauli noise operators are compared for the same model.

Hereafter, we give additional details regarding the numerical simulations.

\subsection{Noiseless data for the critical adiabatic loop in the Ising model}

\begin{figure*}[t!]
\includegraphics[width=1.0\linewidth]{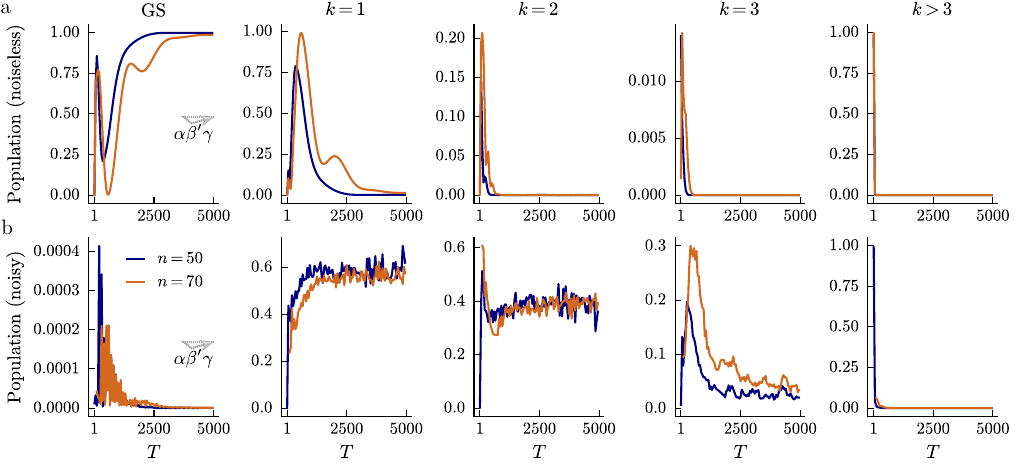}
    \caption{Complementary data to Fig.~\ref{fig:Fig3}(d) for the critical loop $\triangle\alpha\beta'\gamma$. We show the ground state fidelity and the subspace populations for both the noiseless and the noisy dynamics.}
    \label{fig:App_Fig3}
\end{figure*}

We include the ground state fidelity and the subspace populations for a noiseless adiabatic algorithm following the critical loop ($\triangle\alpha\beta'\gamma$) in Fig.˜\ref{fig:Fig3}(d). This is to showcase the regime where the dynamics are sufficiently adiabatic. Note that the dynamics on 70 spins become approximately adiabatic for an evolution time per leg of the triangle of around $T=5000$, even though there is an initial peak where the ground state population in the noiseless case reaches 80\% at time $T=200$ already. This is due to interference of the low energy states when the second small gap is passed, giving rise to an oscillation of the ground state fidelity as a function of the evolution time. The induced phase shift at the avoided level crossing depends on the velocity at the crossing and gives rise to so-called Stückelberg oscillations~\cite{Shevchenko2010Landau}.

\subsection{Details on the antiferromagnetic phase in the Ising model}

\begin{figure}[htb]
    \includegraphics[width=1.0\linewidth]{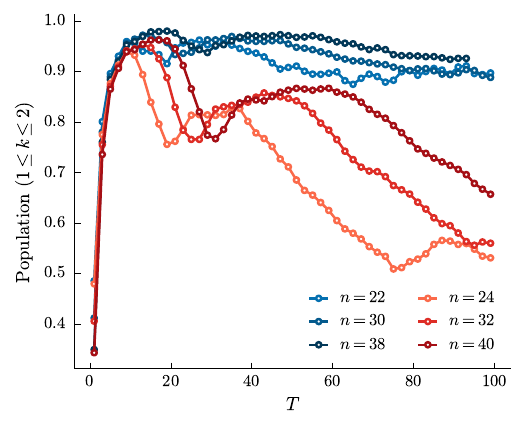}
    \caption{Comparison of the population in the first excitation spaces above the ground state after a quasi-adiabatic loop in the antiferromagnetic phase of the mixed-field Ising model. The blue curves correspond to the case where the position of the erroneous qubit is such that a \ket{\uparrow} gets flipped into a \ket{\downarrow}, and vice-versa for the red lines. Different energy lines get populated for the two cases during the ensuing adiabatic algorithm, resulting in higher or lower final excess energy.}
    \label{fig:App_AFM}
\end{figure}

We comment on a subtlety regarding the proliferation of a single error in the antiferromagnetic phase of the Ising model considered in the main text. In our simulations we apply the error to a center qubit at position $i=n/2$. If the system size is $(n\;\text{mod}\;4) = 0$, the error is applied to a qubit at an even index, for $(n\;\text{mod}\;4) = 2$ the index is odd. The adiabatic path through the antiferromagnetic phase in Fig.~\ref{fig:Fig4}(a) starts at $H_0 = \sum_i \sigma_i^z \sigma_{i+1}^z$, which has a twofold degenerate ground state. We initialize the adiabatic algorithm in the state $\ket{\uparrow_1\downarrow_2\dots\downarrow_{n-1}\uparrow_{n}}$. Then, depending on the system size, the error either acts as $\sigma_i^y \ket{\dots \uparrow_{i-1}\downarrow_i\uparrow_{i+1}}$ or $\sigma_i^y \ket{\dots \downarrow_{i-1}\uparrow_i\downarrow_{i+1}}$. As evidenced by a perturbative argument, under the presence of a field $\sum_i \sigma_i^z$, at the beginning of the adiabatic algorithm, the energy lines split depending on the local longitudinal (and transverse) fields. Then, whether the error effectively flips a up-spin into a down-spin, or vice-verse, leads to the dynamics following larger- or lower-lying energy lines. In Fig.~\ref{fig:Fig4}(b), we observe that the adiabatic algorithm can be relatively robust to a single error in the antiferromagnetic phase. In this case, the population in the first two excitation spaces is approximately $90\%$ and mostly low-lying energy states are populated after the dynamics. However, if the system size is $(n\;\text{mod}\;4) = 0$, then the error leads to overlap with slightly higher energy lines immediately after the error, which in turn, give rise to higher energies after the evolution. For this case, the population in the first two excitation spaces is significantly lower as it can be observed in Fig.~\ref{fig:App_AFM}.

\end{document}